\documentclass[preprint,12pt]{elsarticle}

\usepackage{amssymb}
\journal{Frontiers of Nanoscience}

\begin{document}

\begin{frontmatter}



\title{Free-energy landscapes of DNA and its assemblies: Perspectives from coarse-grained modelling}


\author[PTCL]{Jonathan P. K. Doye\corref{cor1}}
\cortext[cor1]{Corresponding Author}
\address[PTCL]{Physical and Theoretical Chemistry Laboratory, Department of Chemistry, University of Oxford, Oxford, OX1 3QZ, UK}
\author[OxfordPhys]{Ard A. Louis}
\address[OxfordPhys]{Rudolf Peierls Centre for Theoretical Physics, University of Oxford, Parks Road, Oxford, OX1 3PU, UK}
\author[NCAR]{John S. Schreck}
\address[NCAR]{National Center for Atmospheric Research, Computational and Information Systems Laboratory, 850 Table Mesa Drive, Boulder, CO 80305, USA}
\author[Venice]{Flavio Romano}
\address[Venice]{Dipartimento di Scienze Molecolari e Nanosistemi, Università Ca’ Foscari di Venezia, via Torino 155, 30170 Venezia Mestre, Italy}
\author[PTCL]{Ryan M. Harrison}
\author[Bristol]{Majid Mosayebi}
\address[Bristol]{School of Mathematics, University of Bristol, Bristol, BS8 1QU, UK}
\author[Harvard]{Megan C. Engel}
\address[Harvard]{School of Engineering and Applied Sciences, Harvard University, 29 Oxford Street, Cambridge, Massachusetts 02138, USA}
%\author[PTCL]{Chak Kui Wong}
%\address{Physical and Theoretical Chemistry Laboratory, Department of Chemistry, University of Oxford, Oxford, OX1 3QZ, United Kingdom}
\author[Imp]{Thomas E. Ouldridge}
\address[Imp]{Department of Bioengineering and Centre for Synthetic Biology, Imperial College London, London, SW7 2AZ, UK}

\begin{abstract}
This chapter will provide an overview of how characterizing
free-energy landscapes can provide insights into the biophysical properties of
DNA, as well as into the behaviour of the DNA assemblies used in the field of
DNA nanotechnology. The landscapes for these complex systems are accessible through the use of accurate coarse-grained descriptions of DNA. Particular foci will be the landscapes associated with DNA self-assembly and mechanical deformation, where the latter can arise from either externally imposed forces or internal stresses.  
%Particular foci will be the landscapes associated with the pathways and kinetics of self-assembly, and those associated with mechanically-stressed states. In the area of DNA nanotechnology we will consider both nanodevices powered by hybridization and toehold-mediated strand displacement, and the additional complexities that arise when considering theassembly of increasingly large DNA nanostructures
\end{abstract}

%begin{graphicalabstract}



\end{frontmatter}



\section{Introduction}
\label{sect:Intro}

Free-energy landscapes 
depict the free energy of a system as a function of a limited number 
of coordinates (often termed order parameters or reaction coordinates). These landscapes can provide a simplified representation of the main features of configuration space, for example, the (meta)stable states of the system, the pathways between these states in this reduced order-parameter space and the size of the associated free-energy barriers.

These landscapes provide a much more coarse-grained view of a system than a potential-energy surface (PES). For example, the number of potential energy wells that contribute to a particular point on the free-energy landscape may be enormous, particularly for large systems and those where some of the states involve structural disorder. As such the free-energy landscapes naturally capture the effects of entropy and, unlike a PES, free-energy landscapes are specific to the temperature of interest. 
As the projection of the system's configurations onto a limited number of order parameters inevitably leads to a loss of information, the choice of these order parameters is particularly important if the free-energy landscapes are to provide a useful representation of the system; some of the problems associated with bad choices are well documented \cite{Peters16}. Even with a good choice, however, the properties of the landscapes are still intrinsically order-parameter dependent.

\begin{figure}[t]
	\begin{center}
	\includegraphics[width=104mm]{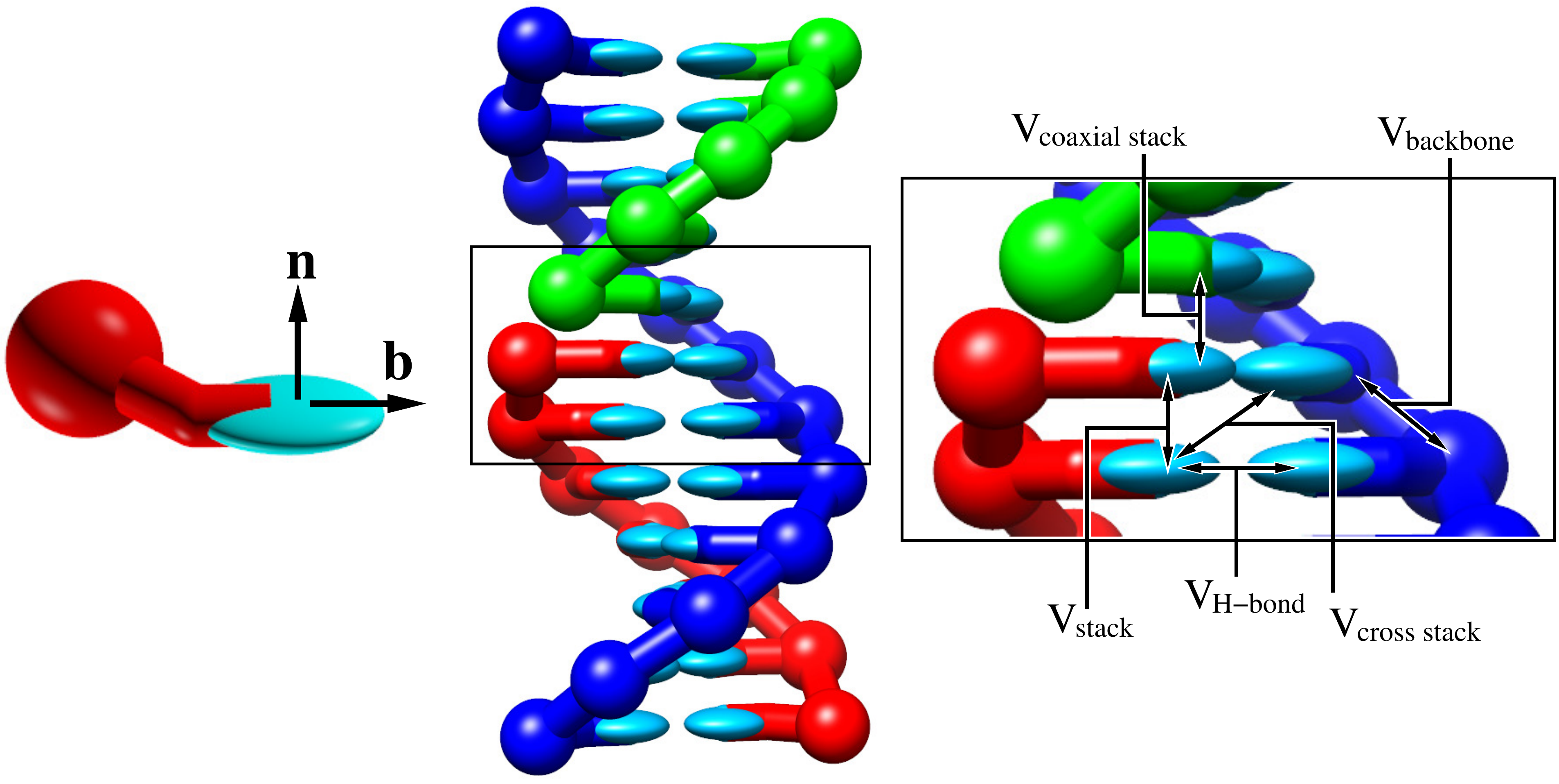}
	\caption{The oxDNA model. Each nucleotide is a rigid body with sites corresponding to the centres of the different interactions. The position and orientation of a nucleotide are defined by the position of the notional centre of mass,  a “base” vector $\mathbf{b}$ that is collinear with the stacking and hydrogen-bonding sites and  a vector $\mathbf{n}$ that is normal to the notional plane of the base. The basic interactions in the model are the backbone potential connecting backbone sites, a hydrogen-bonding potential between complementary nucleotides, (coaxial) stacking interactions between bases that are (non-)adjacent along the chain, electrostatic repulsion between backbone sites, and cross-stacking interactions between bases that are diagonally opposite in the duplex. The images are of the second version of the oxDNA model in which the major and minor grooves have different widths.
	}
\label{fig:oxDNA}
\end{center}
\end{figure}

Here, we will illustrate the utility of analysing systems in terms of their free-energy landscapes for a series of systems made out of DNA. As the systems of interest would often involve an extremely large number of atoms if all were explicitly represented, we instead represent the DNA using a coarse-grained model of DNA at the nucleotide level, thus allowing us to access much larger systems and much longer time scales. Specifically, we use the oxDNA model \cite{Ouldridge11,Sulc12,Snodin15}, which models the DNA as a set of rigid nucleotides with a set of interactions representing the bonding along the backbone, Watson-Crick base-pairing, stacking, coaxial stacking, cross-stacking, and electrostatics, as illustrated in Fig. \ref{fig:oxDNA}. 
One of the dangers of coarse-graining is that the removed degrees of freedom are in some way coupled to the process of interest and cannot simply be ``integrated out''. In the case of oxDNA, it is probably reasonable to assume that the intramolecular degrees of freedom of the nucleotide do not change significantly in most of the processes in which we are interested, but the potential effects of only including the solvent degrees of freedom implicitly are somewhat less clear.

In the development of the oxDNA model a particular focus was on the accurate representation of the thermodynamics of hybridization and a good description of single-stranded DNA and double-stranded DNA mechanics (i.e.\ the thermodynamic cost of structural deformations). This makes the model particularly well-suited for the computation of free-energy landscapes associated with the biophysical properties of DNA and for systems of interest to DNA nanotechnology. Although certain features of these landscapes are relatively straightforward to predict from simpler models (e.g.\ the Santa Lucia model for hybridization free energies \cite{SantaLucia04} and the worm-like chain model for DNA mechanics \cite{Brahmachari18b}), the oxDNA model naturally captures the interplay of configurational entropy (particularly of single-stranded sections), geometric effects resulting from the double-helical structure of double-stranded DNA and DNA mechanics that go into determining the size of free-energy barriers for intramolecular processes, and which would be otherwise hard to estimate accurately. 

We should note that the oxDNA model  has a number of variants. In particular a second version of the model (sometimes called oxDNA2) was developed that introduced explicit electrostatic interactions and fine-tuned structural parameters to enable the structure of large DNA nanostructures, such as DNA origami, to be accurately described \cite{Snodin15,Snodin19}. Although this second model is now generally the version of choice, the predictions of the two models will be very similar for many systems, and the results reported here will be for both models. In addition, one has the option to use a sequence-averaged or sequence-dependent \cite{Sulc12} parameterization, where in the former the strength of the interactions are independent of the identity of the nucleotides involved. Most of the results reported here are for the sequence-averaged parameterization, as it allows the generic properties of systems to be elucidated, whereas the sequence-dependent parameterization tends to be reserved for detailed comparisons to specific experimental systems.

\section{Computing free-energy landscapes}

Free-energy landscapes are simply related to equilibrium probability distributions through
\begin{equation}
    A(\{Q_i\})=-k_B T \log p(\{Q_i\}) + c
\end{equation}
where $\{Q_i\}$ is a set of order parameters that are being used to characterize the system and $c$ is an arbitary constant. To facilitate visualization of the free-energy landscape often only one or two order parameters are used. 

The major difficulty in computing such free-energy landscapes is that one is often interested in high free-energy regions of the landscape --- say because one is interested in the size of the free-energy barrier for a process --- which have a very low probability and so will be poorly sampled in standard simulations. To overcome this problem, there is now a large array of rare-event and enhanced sampling simulation methods that allow free-energy landscapes or the rates of slow processes on such landscapes to be computed \cite{Kastner11,Allen2009,Bolhuis15,Valsson16}.
In our work on the free-energy landscapes of DNA systems, we have almost exclusively used one of the simplest and earliest of such methods, namely umbrella sampling \cite{Kastner11,Torrie77}. This involves adding an additional biasing potential $V_\mathrm{bias}(\{Q_i\})$ 
to the potential energy in order to improve the sampling of the higher free-energy regions of the landscape. From the probability distribution in a biased simulation $p_\mathrm{bias}(\{Q_i\})$ it is a simple matter to estimate the same distribution for the unbiased system.

There can be different ways to implement umbrella sampling. One limit is to choose the biasing potential so that all regions of the landscape can be well sampled. This has the potential disadvantage that the simulation has to be sufficiently long that the system is able to diffuse back and forth across the whole order parameter space a sufficient number of times for the relative free energies of different parts of the landscape to be correctly estimated. A second issue is generating the biasing potential. Ideally, one would want $p_\mathrm{bias}(\{Q_i\})$ to be flat, but to achieve this would require $V_\mathrm{bias}(\{Q_i\})=-A(\{Q_i\})$, i.e.\ one would need to know the quantity that one is seeking to calculate. One typically gets around this problem by running a series of simulations in which the biasing potential is iteratively improved until sampling is sufficiently uniform.

The other limit is to run many separate simulations with different biasing potentials, the aim being that each characterizes a different region of the landscape. One simple way to achieve this is to use harmonic biasing potentials whose minima form a grid in the order-parameter space. A best estimate of the whole free-energy landscape can then be constructed from the set of biased probability distributions using the weighted-histogram analysis method (WHAM) \cite{Kumar92}. For this to be accurate it is important that there is sufficient overlap between the probability distributions for adjacent points on the order-parameter grid. 

This approach has a number of potential advantages. Firstly, it is potentially much more efficient, simply because the simulation time required to diffuse over an order parameter window $\Delta Q$ is expected to be proportional to $(\Delta Q)^2$. Thus, this analysis would predict for one-dimensional umbrella sampling that the total time required to accurately sample the landscape when using 
$N_w$ windows is a factor of 
$1/N_w$ shorter compared to a simulation with one window \cite{Chandler}. Secondly, this approach naturally lends itself to trivial parallelization through task farming. 

However, the major disadvantage is that it can be much easier for problems with the sampling to go unnoticed. In particular, it is often the case that even when the system is at the top of a free-energy barrier transitioning between states that are representative of the two basins that meet at that point may still be an activated process and hence slow. A good estimate of the relative free-energy of the two basins can only be obtained if for the windows in the transition region the systems transits back and forth between states associated with the two basins sufficiently frequently. 

Typically, we use an intermediate approach with a limited number of overlapping windows; those that cover a transition region often need to be run for significantly longer.

\section{Example free-energy landscapes}

\subsection{Hybridization}

Hybridization, the association of two strands to form a double helix, represents the most fundamental process of DNA self-assembly. Consequently, this process has been well characterized experimentally with the thermodynamics well understood \cite{SantaLucia04,Dirks07} but with open questions still remaining concerning the hybridization kinetics \cite{Ouldridge13b,Andrews21}.

\begin{figure}[t]
	\begin{center}
	\includegraphics[width=104mm]{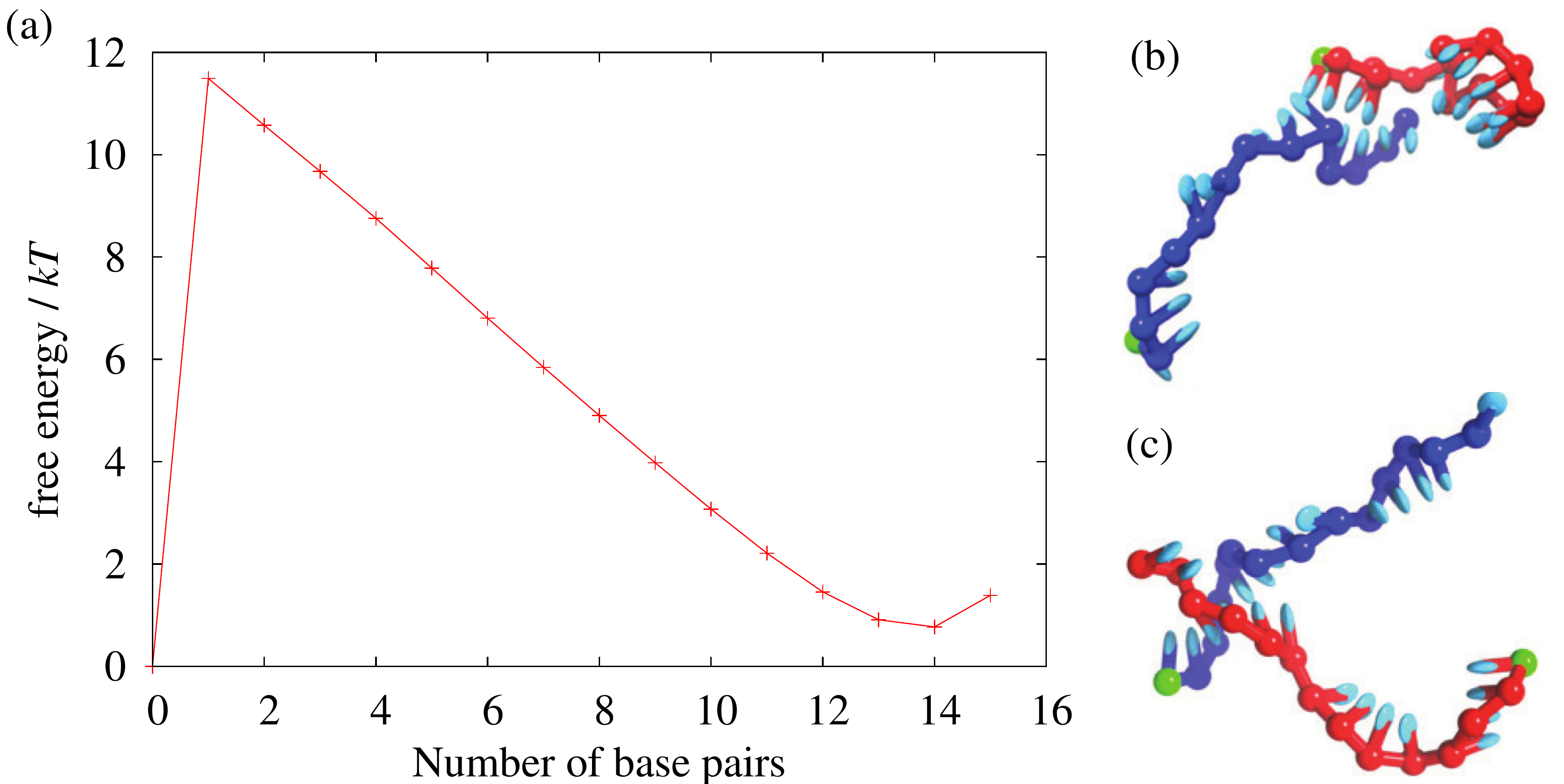}
	\caption{
	(a) Free-energy profile for the hybridization of a 15-bp duplex as a function of the number of base pairs formed at 343\,K and a strand concentration of $3.36\times 10^{-4}$\,M \cite{Ouldridge11}. (b,c) Example isoenthalpic configurations with 2 base pairs formed from (b) an association simulation and (c) an equilibrium simulation. The configuration in (b) requires more extensive structural rearrangements of the single-stranded tails to allow further base pairs to form \cite{Ouldridge13b}.
	}
\label{fig:hybrid}
\end{center}
\end{figure}

The most basic measure of the degree of assembly is the number of base pairs formed. The free-energy profile as a function of this order parameter is shown in Fig.\ \ref{fig:hybrid}(a). The main features of this landscape 
are (i) the barrier associated with the formation of the first base pair due to the loss of translational
entropy, (ii) the roughly linear down hill slope as the duplex zips up and (iii) deviations from linearity for the final base pairs due to the propensity of the end base pairs to ``fray''. 

Variants of this basic landscape, often as components of a larger landscape, will be seen a number of times in this review. However, it is important to be aware of a number of additional complexities. Firstly, when picking a reaction coordinate to compute a free-energy profile one usually wants to choose the slowest degree of freedom in the system. If that is the case, motions orthogonal to that coordinate would be expected to be in equilibrium as one dynamically progresses along the reaction coordinate. In the case of hybridization, though, structural rearrangement of the unhybridized sections of the chain is another slow coordinate, leading to dynamical effects not expected from the landscape. For example, even though a state with one base pair is the free-energy transition state on assembly, two to three base pairs must form before the probability of hybridization exceeds 50\%. This is because when one base pair is formed the rearrangements of the chain required to form the next base pair are slow (Fig.\ \ref{fig:hybrid}(b)), with breaking of the base pair occurring at a faster rate. Similarly, on melting, states with one base pair are more likely to zip up than dissociate because the chains still adopt conformations that allow base pairing to occur relatively easily. 

Secondly, the implicit description of the solvent in coarse-grained models such as oxDNA may neglect important degrees of freedom for the hybridization process. For example, the oxDNA barrier to hybridization is purely entropic and the oxDNA transition state is enthalpically stabilized with respect to the single-stranded reactants because of the base pair(s) formed at the transition state. Hence, the oxDNA hybridization rate decreases with temperature because of this negative activation energy \cite{Ouldridge13b}. However, in reality, hybridization rates generally increase with temperature \cite{Zhang18b}. Thus, there must be an enthalpic barrier to hybridization that is not captured by oxDNA. The most likely source is the water degrees of freedom, for example, desolvating the bases and the resulting rearrangements of the hydrogen-bond network.

What have we learnt from this free-energy landscape? Not that much in some ways. The loss of translational entropy that is the dominant contributor to the free energy barrier can be easily estimated from standard statistical mechanics 
and, as the oxDNA model has been parameterized to the predictions of the SantaLucia model, the free-energy of binding should be very similar to that model. The most non-trivial feature is perhaps the change in slope due to fraying; this has not been ``put into'' oxDNA, but emerges naturally from its relatively realistic description of the basic physical properties of the nucleotides. 

\subsection{Hairpin formation}

\begin{figure}[t]
	\begin{center}
	\includegraphics[width=104mm,angle=0]{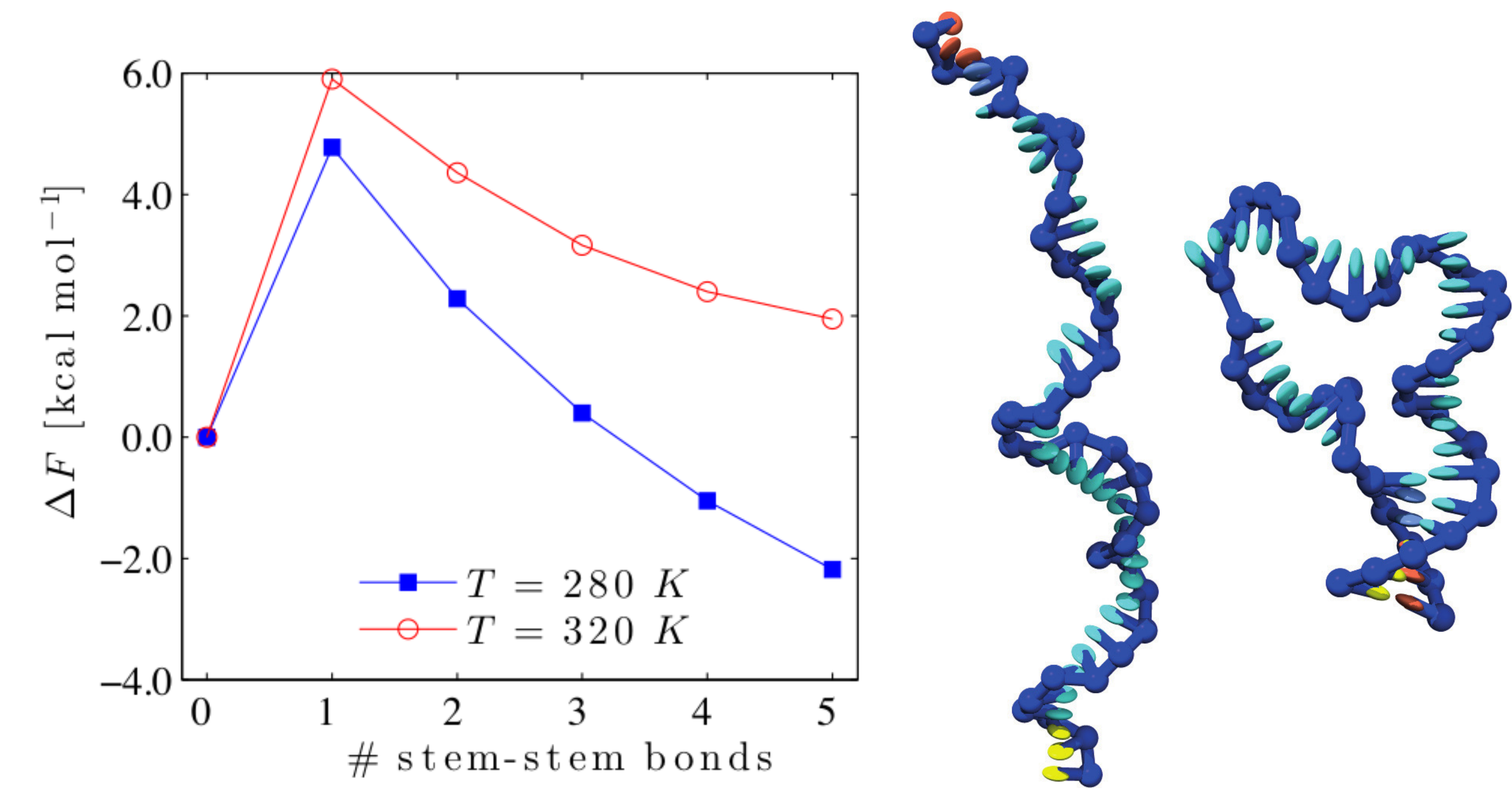}
	\caption{
	Free-energy profile for the formation of a hairpin with a 5-bp stem and a 30-nt loop \cite{Mosayebi14}. Example open and closed configurations are also illustrated.
	}
\label{fig:hairpin}
\end{center}
\end{figure}

The next example that we consider is the formation of a hairpin. The free-energy profile as a function of the number of stem base pairs formed is shown in Fig.\ \ref{fig:hairpin} at two temperatures. Its form is very similar to that for hybridization, but for this unimolecular process the barrier is instead a result of the loss in configurational entropy on forming the hairpin loop. One of the particular strengths of the oxDNA model is its ability to capture such features, particularly as describing single-stranded DNA accurately using polymer models is not so straightforward. This enables the model to reproduce hairpin melting temperatures accurately, albeit noting that the reproduction of such data was considered in the model parameterization. 

\subsection{Kissing hairpins}

\begin{figure}[t]
	\begin{center}
	\includegraphics[width=104mm,angle=0]{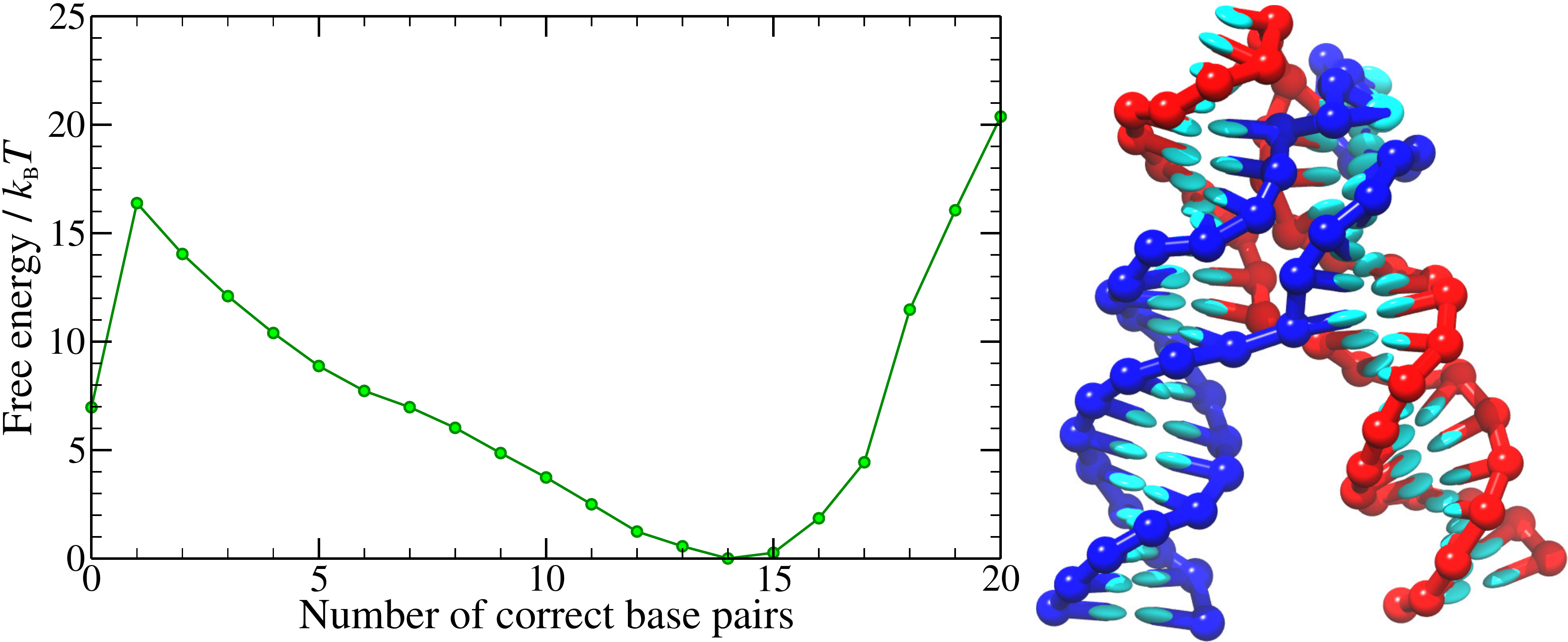}
	\caption{
	The free-energy profile for the formation of a kissing-loop between two hairpins with fully complementary 20-nt loops at 23$^\circ$C and a strand concentration of 0.336\,mM \cite{Romano12b}. The illustrated configuration with 14 base pairs formed between the loops is representative of the free-energy global minimum.
	}
\label{fig:kissing}
\end{center}
\end{figure}

The formation of the DNA double helix requires the wrapping of each strand around each other in a right-handed fashion. In some instances, this wrapping may not be topologically feasible. Fig.\ \ref{fig:kissing} illustrates one such example, namely the hybridization of two fully complementary hairpin loops (such a structure is termed a kissing loop). As the loops are longer than the DNA pitch length and there are no free ends, the two loops cannot fully wrap around each other without defects associated with the the opposite wrapping. The topological and geometric frustration in this system is evident from the free-energy profile. Although the formation of the first few base pairs is relatively unhindered, the graph shows increasing curvature as more base pairs are formed, and the lowest free-energy state has only 14 of the 20 possible base pairs formed. 

Somewhat similar topological impediments to assembly may well be relevant to DNA origamis. For example, if both the end domains of a staple strand bind first to the scaffold strand, then any remaining unbound middle domains no longer have a free end to facilitate wrapping and so correct binding of that domain may no longer be topological feasible without dissociation of one of the end domains.

\subsection{Force-induced melting}

\begin{figure}[t]
	\begin{center}
	\includegraphics[width=144mm,angle=0]{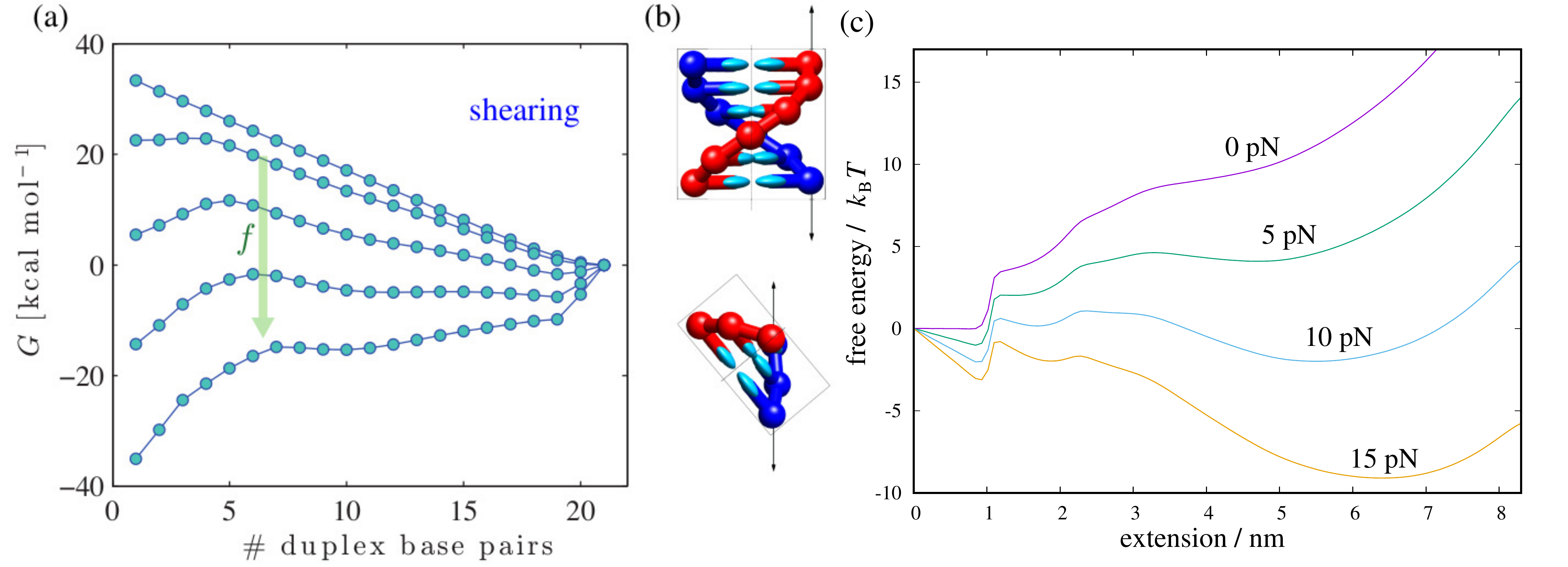}
	\caption{
	(a) Free-energy profiles for force-induced melting of a 20-bp duplex when subjected to tensile forces applied in a ``shearing'' mode of 0, 30, 50, 70, 90\,pN \cite{Mosayebi15}. (b) The configurations illustrate the changes in orientation of the double-stranded section when the number of base pairs is less than a pitch length. 
	(c) Free-energy profile for the force-induced ``unzipping'' of a hairpin with a 6-bp stem and 
	a 4-nt loop as a function of the extension (the end-to-end distance of the strand along the direction of force) at a temperature of 30$^\circ$C and at four different forces as labelled.
	}
%\label{fig:shearing}
\label{fig:force}
\end{center}
\end{figure}

The hairpin free-energy profiles in Fig.\ \ref{fig:hairpin} illustrate the destabilization of the assembled state by temperature. 
Another mechanism of destabilization is by the application of stress to the DNA systems. Fig.\ \ref{fig:force}(a) illustrates the destabilization of a DNA duplex under the action of tensile force applied in a `shearing' mode (the forces are applied at opposite ends of the duplex on different strands). As the force $F$ increases, the duplex becomes increasingly destabilized because there is now an additional $-F\Delta z$ contribution to the free energy of breaking a base pair, where $\Delta z$ is the extension gained in the direction of the force from breaking a base pair. Beyond 10 base pairs (approximately the pitch of DNA) the effect of the force is mainly just to change the slope of the profile, but below this value of the reaction coordinate the profile develops greater structure. 
In this limit, the orientation of the intact double-helical section varies with the number of remaining base pairs (as illustrated by the configurations in Fig.\ \ref{fig:force}(b)) and hence so does the extension gained when a base-pair breaks. 

Of course, the number of base pairs is not the only reaction co-ordinate that one can potentially use. In the case of force-induced transitions, the free-energy profile as a function of the extension provides a representation that one can more directly relate to experiments using optical tweezers \cite{Woodside06}. Such landscapes are illustrated in Fig.\ \ref{fig:force}(c) for a hairpin with a 6-bp stem where tensile forces are applied to both end nucleotides. 
The free-energy minimum at 1\,nm corresponds to the oriented closed state (1\,nm is the separation of the oxDNA nucleotide centres in a base pair), whereas the minimum at $\sim$5-7\,nm that appears at higher forces corresponds to the open state. The metastable minimum at 2\,nm corresponds to a hairpin with the end base pair broken, but no further intermediate sub-states are resolvable. The stabilization of the open state with increasing force is clear, and the centre of the transition from the closed to the open state occurs at close to 10\,pN.

\subsection{Duplex bending}

\begin{figure}[t]
	\begin{center}
	\includegraphics[width=124mm,angle=0]{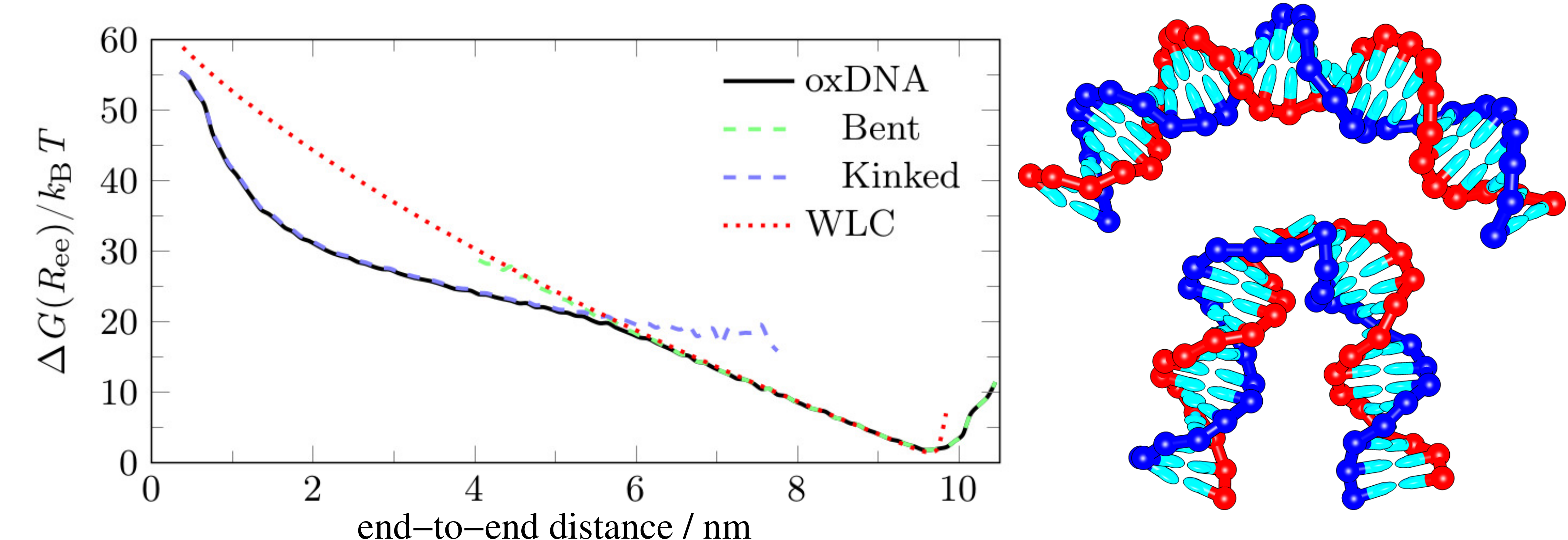}
	\caption{
	Free-energy profile for the bending of a 30-bp DNA duplex using the end-to-end distance as the order parameter. The contributions to the free-energy from homogeneously bent and kinked states are included on the plot; example configurations for these two types of states are depicted. A worm-like chain fit to the free-energy of the homogeneously bent duplex is also shown \cite{Becker10}.
	}
\label{fig:bend}
\end{center}
\end{figure}

Free-energy landscapes can also be computed that quantify the effect of mechanical deformation on double-stranded DNA. In Fig.\ \ref{fig:bend} we show the thermodynamic cost of duplex bending, using the end-to-end distance of the duplex as the order parameter. The free-energy minimum corresponds to the relaxed duplex. As the two ends of the duplex are pulled together the free-energy rises smoothly in line with the predictions of the worm-like chain model. However, below a certain end-to-end distance (about 5.5\,nm in the example shown) it become favourable for the bending stress to be localized at a ``kink'' defect where the DNA bends sharply and the stacking and usually base-pairing interactions are disrupted. The kink acts a bit like a hinge allowing further bending to occur more easily. At very short distances, the free energy begins to rise more rapidly due the repulsion between the two ends. The point at which kinking occurs in such free-energy profiles depends on the duplex length; for sufficiently long duplexes the two ends can be brought together to form a roughly circular configuration without any need for kinking \cite{Harrison15}.

\subsection{DNA nanodevices}

\begin{figure}
	\begin{center}
	\includegraphics[width=134mm,angle=0]{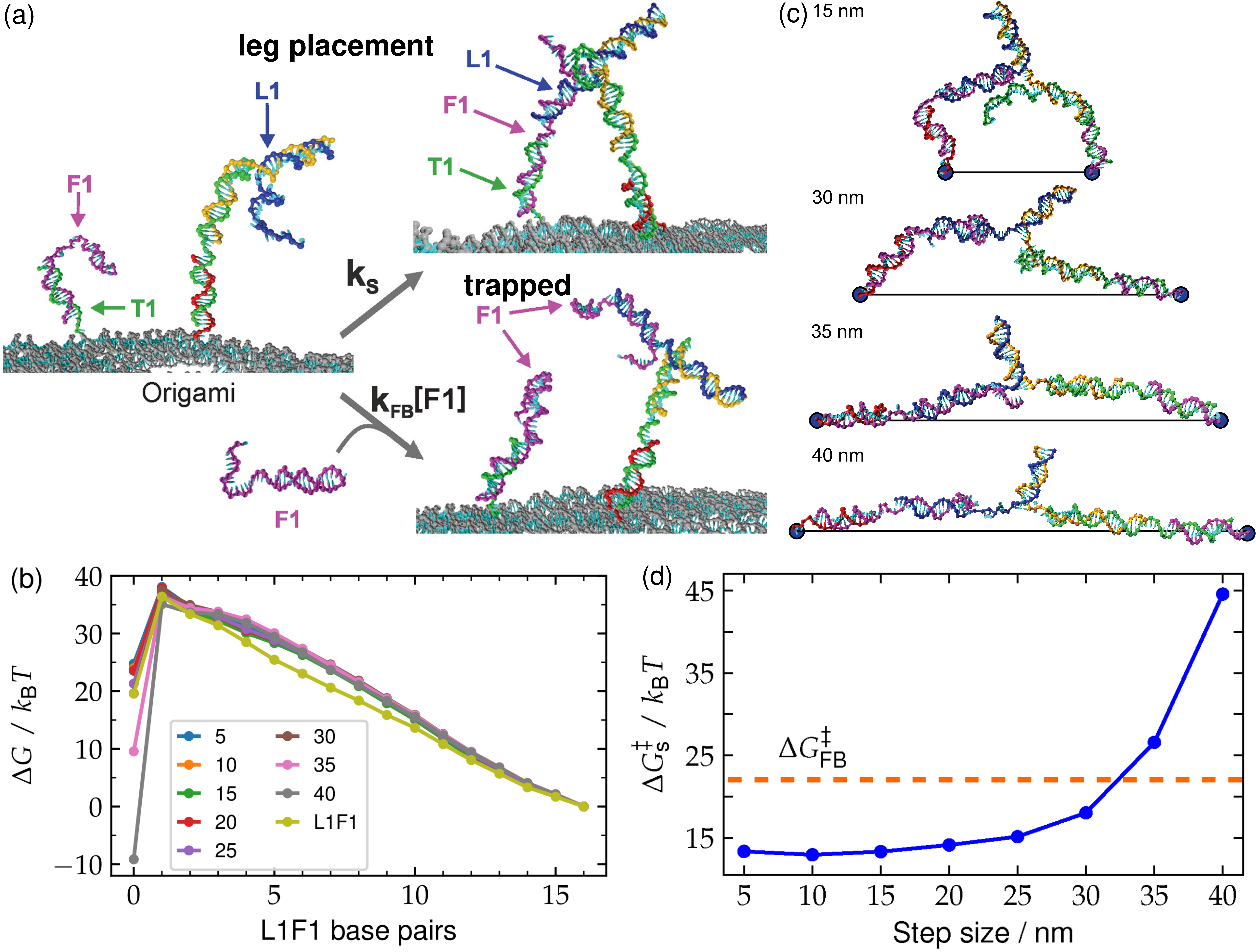}
	\caption{(a) The leg placement step of a two-legged DNA walker that is capable of walking between footholds on a DNA origami, together with a competing reaction involving the binding of an additional fuel strand F1 that leads to a trapped state. (b) Free-energy profiles for the binding of L1 to F1 as a function of the step size (the separation between the footholds on the DNA origami) together with that for the trapping reaction.  (c) Example configurations with the foot fully bound at different step sizes. (d) The free-energy barrier for foot placement as a function of step size, compared to that for the trapping reaction (red dashed line) at a F1 concentration of 10\,nM.
	}
\label{fig:walker}
\end{center}
\end{figure}

In many DNA nanodevices, part of the mechanism  involves an intramolecular hybridization process. The two-footed DNA walker of Ref.\ \cite{Tomov13} provides just such an example. This device is capable of walking across a DNA origami through the successive additions of fuel and anti-fuel molecules that cause the legs to be lifted and placed back down. The process illustrated in Fig.\ \ref{fig:walker}(a) is of leg placement after a fuel molecule F1 has bound to the foothold T1. The intramolecular hybridization of the leg L1 with the remaining unbound domain of F1 leads to correct foot placement, whereas the binding of a second F1 molecule to L1 instead leads to a trapped state.

Fig.\ \ref{fig:walker}(b) shows the free-energy landscapes for leg placement as a function of the step size of the walker. Beyond a step size of 25--30\,nm, the free-energy for leg placement increases rapidly. This is because there is an increasing free-energetic cost for the walker adopting a stretched-out state, as is very evident from the bound states illustrated in Fig.\ \ref{fig:walker}(c). 
In terms of free-energy barriers, the main effect of increasing the step size is to increase the barrier for binding (Fig.\ \ref{fig:walker}(d)); most of the free-energy cost of adopting a stretched-out state has to be paid to initiate binding. Thus, the step size has a much greater effect on the binding rate than the unbinding rate.

Although estimating absolute rates from free-energy landscapes requires additional dynamical information, the relative rates for similar processes can often be accurately estimated by assuming that dynamical factors other than the free-energy barriers cancel out; i.e.\
\begin{equation}
\frac{r_1}{r_2}\approx\exp\left(-\frac{(\Delta G^\ddagger_1-\Delta G^\ddagger_2)}{k_B T}\right)
\label{eq:relrate}
\end{equation}
where $r_1$ and $r_2$ are the rates of the two reactions being compared. 
For example, this equation allowed a detailed comparison to experimental results for the walker,  considering both the effects of step size and fuel concentration \cite{Khara18}. 

The above provides an example where a change in the free energy of reaction mainly affects the binding rates. This is often the case when a mechanical deformation is required to initiate binding; for example, this also applies to DNA cyclization reactions \cite{Harrison19}.
An opposite example is the hybridization rate between two strands that can themselves form short hairpins \cite{Gao06,Schreck15b}.
The hairpins stabilize the single-stranded state compared to sequences that do not allow hairpins to form, thus reducing the overall free energy of hybridization. However, the hairpins only have a relatively small effect on the binding rate because, in the cases considered, the hairpins are short enough that hybridization can be initiated in other sections of the strand, and only once the free-energy barrier has been crossed do the hairpin base pairs need to be broken. By contrast, there is a large increase in the rate of unbinding, because during dissociation the strands can start to form the hairpins, thus stabilising partially melted states and reducing the free-energy barrier to dissociation.

\subsection{Displacement reactions}

\begin{figure}[t]
	\begin{center}
	\includegraphics[width=124mm,angle=0]{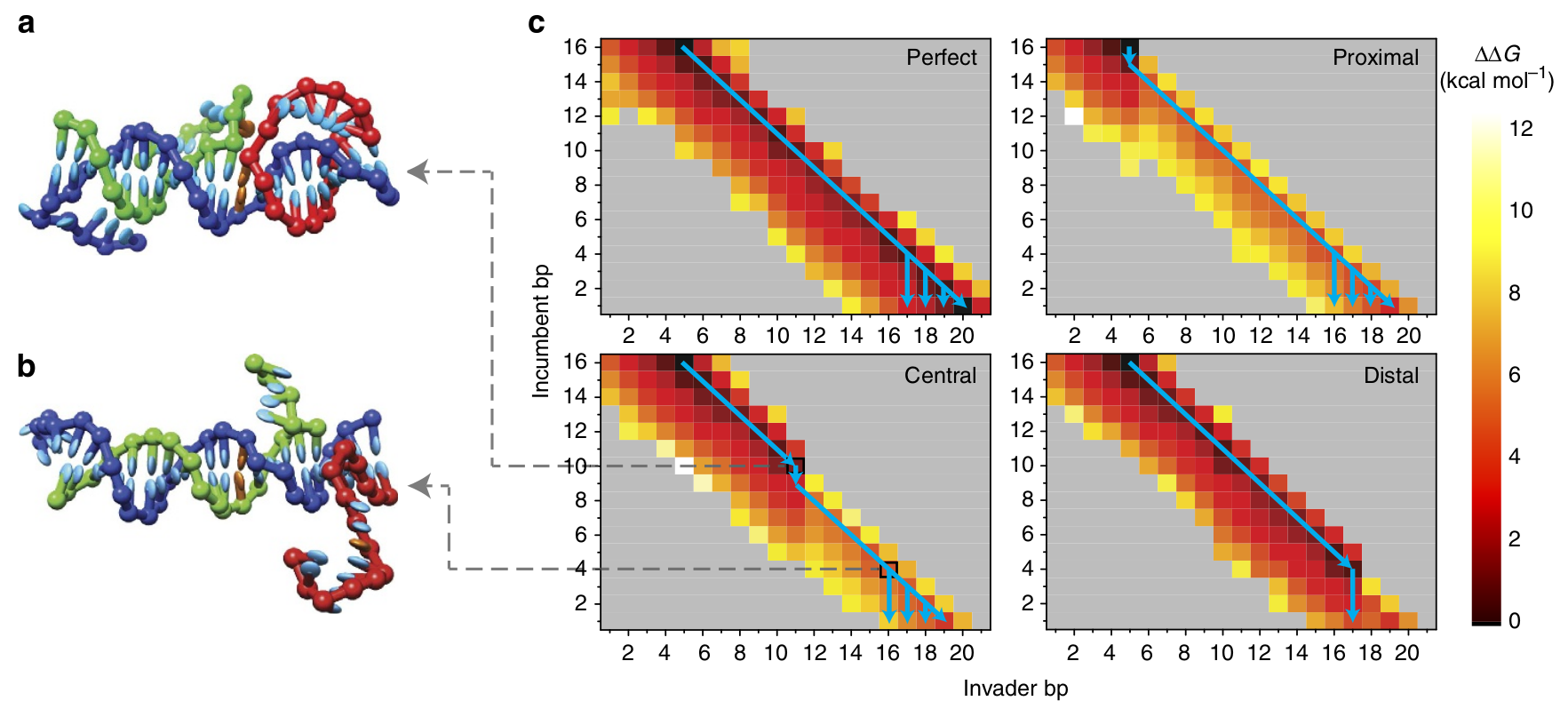}
	\caption{Toehold-mediated strand displacement involving the creation of a mismatch. (a,b) Example configurations from different points in the displacement reaction where the green invader strand displaces the red incumbent strand from the blue target strand \cite{Machinek14}. The gold bases indicate the site where a mismatch is created, and in (a) the invader is bound up to the mismatch site, whereas in (b) the invader has passed beyond the mismatch.
	(c) Free-energy landscapes for four different invading strands: the fully complementary invader, and invaders creating mismatches proximal to the start of the displacement, in the centre or at the distal end. The toehold is of length 5, and so state (5,16) corresponds to complete binding of the invader to the toehold but without any displacement. The blue lines correspond to the paths of displacement.
	}
\label{fig:displacement}
\end{center}
\end{figure}

One of the fundamental processes used in DNA nanotechnology to drive structural changes is toehold-mediated strand displacement. In this process an ``invader'' strand binds to an exposed single-stranded toehold on the ``target'' strand and then proceeds to displace the adjacent ``incumbent'' strand to form a duplex with the target strand (example intermediate configurations are shown in Fig.\ \ref{fig:displacement}(a,b)). The first panel of Fig.\ \ref{fig:displacement}(c) shows the landscape for a standard displacement reaction where the invader is fully complementary to the target. The pathway involves the invader binding to the toehold and then the displacement process, which at each step involves the loss of a base pair by the incumbent followed by the invader gaining a base pair (leading to the diagonal paths across the landscape). The final stage is dissociation of the incumbent strand, which may occur spontaneously before the last
few base pairs have been displaced.

A key property that determines the rate of displacement is the likelihood that the invading strand will dissociate rather than complete the displacement. For example, the rate increases as the toehold length increases, because it makes dissociation less likely once the invading strand is fully bound to the toehold. One of the features of the displacement reaction is that there is a small barrier (of about $3k_\mathrm{B}T$) to initiating the displacement, as this then creates two single-stranded tails at the junction site that have an effective repulsion between them. Accounting for this barrier is important to capture the dependence of the displacement rate on toehold length (which oxDNA very accurately does) because it increases the probability of invader dissociation compared to a landscape where there is no initial barrier to displacement \cite{Srinivas13}.

In Fig.\ \ref{fig:displacement}(c) we also show landscapes for sequences where the displacement creates a duplex that contains a mismatch \cite{Machinek14}. Incorporation of this mismatch leads to an additional barrier that needs to be overcome for the displacement reaction to reach completion. When incorporation of the mismatch occurs early in the displacement process, close to the toehold, the rate of displacement is significantly reduced (by a factor of about 1000 compared to the fully complementary invader sequence) because this additional barrier increases the likelihood that the invader strand will dissociate from the target strand rather than surmount the barrier and complete displacement. By contrast, when the mismatch is at the distal end relative to the toehold, it only has a relatively small effect on the displacement rate (a reduction by a factor of roughly 2), because by that point the invading strand is likely to complete displacement anyway, rather than dissociate, because it has already formed 17 base pairs with the target. The use and positioning of such mismatches thus provides an additional means to control the dynamics of displacement reactions.

\section{Conclusions}

This chapter provides some examples of how free-energy landscapes computed using a coarse-grained DNA model can provide important insights into the biophysics of DNA and the properties of the DNA assemblies used in DNA nanotechnology. Furthermore, given that the model used (oxDNA) has been parameterized to reproduce accurately the thermodynamics and mechanics of DNA both in its double-stranded and single-stranded forms, these landscapes can provide quantitative insights into relative rates. Such predictions can be supplemented by direct calculations of the rates by simulations (we have generally used the forward flux-sampling approach \cite{Allen2009}; note that if one knows the free-energy change for the reaction from the free-energy landscape and one has computed the rate in one direction, the rate in the other direction simply follows from these two quantities). Although these rate calculations have revealed interesting and sometimes subtle dynamical effects \cite{Ouldridge13b,Schreck15b,Kocar16}, they have also confirmed the utility of Eq.\ \ref{eq:relrate} in capturing the most substantial effects on the relative rates.

Although the illustrative examples shown here are for relatively small systems, it is feasible to compute the full assembly landscapes including all relevant pathways for small DNA nanostructures, such as the DNA tetrahedron developed by the Turberfield group \cite{Goodman05}. For larger nanostructures, such as DNA origami \cite{Rothemund06} and DNA brick systems \cite{Ke12}, such computations are not feasible. However, the landscapes for key or archetypal processes in the assembly of these objects can be computed; for example, these can be used to parameterize kinetic models of assembly at the domain level, as has been done for 2D DNA brick systems \cite{Fonseca18}.

Free-energy landscapes associated with mechanical deformations of very large DNA systems are also feasible using the oxDNA model. For example, free-energy landscapes for the bending of DNA origami rods that are equivalent to Fig.\ \ref{fig:bend}(a) have now been computed using umbrella sampling \cite{Wong21}. Similarly, metadynamics has also been successfully applied to study the mechanics of DNA origami \cite{Kaufhold21}.





%\begin{thebibliography}{00}


%\bibitem{}

%\end{thebibliography}
\end{document}